\documentstyle[aps]{revtex}  
 
\begin{document}

\title{First Principles Study of Structural,
Electronic and Magnetic Interplay in Ferroelectromagnetic Yttrium Manganite}

\author{Alessio Filippetti and Nicola A. Hill}

\address{Materials Department, University of California\\
Santa Barbara, CA 93106-5050\\
Tel. (805) 893-7920\\
Fax: (805) 893-7221\\
e-mail: nahill@mrl.ucsb.edu}

\date{\today}

\maketitle

\begin{abstract}

We present results of local spin density approximation (LSDA)
pseudopotential calculations for the ferroelectromagnet,
yttrium manganite (YMnO$_3$). The origin of the differences between 
ferroelectric and non-ferroelectric perovskite 
manganites is determined by 
comparing the calculated properties of yttrium manganite
in its ferroelectric hexagonal and non-ferroelectric orthorhombic
phases. In addition, orthorhombic YMnO$_3$ is compared with
the prototypical non-ferroelectric manganite,
lanthanum manganite.
We show that, while the octahedral crystal field splitting of the 
cubic perovskite structure causes a centro-symmetric Jahn-Teller
distortion around the Mn$^{3+}$ ion, the markedly different 
splitting in hexagonal perovskites creates an electronic
configuration consistent with ferroelectric distortion. We explain
the nature of the distortion, and show that a local magnetic moment
on the Mn$^{3+}$ ion is a requirement for it to occur.

\end{abstract}

\vspace*{24pt}
{\bf pacs numbers} 75.50.Dd,71.20.-b,77.84.-s,71.15.Mb


\input psfig

\section{Introduction}

The observation of colossal magnetoresistance (CMR) in 
(La,Ca)MnO$_3$\cite{Jin} has prompted a flurry of recent
research on this material and related perovskite structure 
manganites\cite{example}.
The majority of the recent research has focused on manganites in which the
large (A-site) cation is a rare earth from the left hand side of the
lanthanide series. 
The manganites of these large rare earth ions (lanthanum manganite 
through dysprosium manganite) all
crystallize in the cubic perovskite structure (Figure~\ref{perovskite}), 
with the same low
temperature orthorhombic distortion and A-type antiferromagnetic
ordering of the Mn$^{3+}$ ions.

During the recent studies,
many rare earth manganites have been found to show strong
coupling between their magnetic  and structural, or magnetic and
electric order parameters. For example,
a magnetically induced structural phase transition has been observed in
La$_{0.83}$Sr$_{0.17}$MnO$_3$\cite{Asamitsu} indicating strong coupling
between the local magnetic spin moments and the lattice structure.
In Nd$_{0.5}$Sr$_{0.5}$MnO$_3$, strong coupling
between the magnetic spin moments and the electronic charge
carriers was demonstrated when an electronic metal-insulator transition
was induced by an external magnetic field\cite{Kuwahara}.
Indeed the large change in conductivity with applied magnetic field,
which causes colossal magnetoresistance
is believed to originate from a similar type of phase
transition\cite{Kuwahara2}.

The strong coupling between magnetic and electric order parameters
is of particular interest in the manganites of yttrium and the
smaller (holmium through lutetium) rare earths at the right hand
side of the lanthanide series.  In these materials, the
hexagonal perovskite structure (Figure~\ref{hexperovskite}) is the
lowest energy structure. However
the metastable orthorhombic cubic perovskite phase 
can be regained by annealing at high pressure\cite{Wood}, 
or by appropriate choice of synthetic method. 
The phase transition between the stable hexagonal phase and
the metastable orthorhombic phase is reconstructive and
first order.

In the hexagonal manganites, as in the cubic manganites,
superexchange between adjacent Mn$^{3+}$ ions causes antiferromagnetic
ordering.
However in addition to the magnetic ordering,
the hexagonal manganites also undergo a phase transition
to a non-centrosymmetric low temperature state which has
{ \it ferroelectric} ordering.
Therefore the hexagonal yttrium and rare-earth manganites belong
to the rather small group of materials known as ferroelectromagnets,
in which magnetic and ferroelectric ordering occurs simultaneously.
Although there are thirteen point groups which permit the occurrence
of spontaneous magnetization and spontaneous polarization in
the same phase\cite{Schmid}, remarkably few ferroelectric materials
show any kind (ferro- ferri- antiferro-, etc.) of magnetic ordering. 
Understanding
the fundamental physics that tends to discourage simultaneous
ferroelectricity and magnetic ordering is an area of current research
activity\cite{HillJPC}. And
the fact that the ferroelectric hexagonal manganites can
be readily transformed to their non-ferroelectric cubic form makes
them ideal prototypes for studying this
phenomenon.

The goal of the study described in this paper is to explore theoretically
the similarities and differences between non-ferroelectric cubic  
manganites and their ferroelectric hexagonal counterparts.
To achieve this goal we evaluate the electronic
and magnetic properties of some representative
perovskite manganites using a plane
wave pseudopotential (PWPP) implementation of density functional theory
(DFT) within the local spin density approximation (LSDA).
Our choice of theoretical approach is influenced by
successful first-principles studies of {\it non-magnetic} perovskite
ferroelectrics\cite{Waghmare_Rabe,Cohen_Krakauer}, and by earlier
work which elucidated the origins of the differences between
related cubic perovskite manganites\cite{HillRabe,Pickett_Singh}.
We calculate the differences between the electronic properties
of cubic and hexagonal YMnO$_3$ to determine the
reasons for the lack of ferroelectricity in cubic manganites, and
its presence in hexagonal manganites.
We also compare our calculated results for
cubic YMnO$_3$ with those for the prototypical cubic perovskite
manganite, LaMnO$_3$, to establish whether the  
metastability of the {\it cubic} phase of YMnO$_3$
is merely the result of the small size of the
yttrium ion, or if electronic factors also play a role. 

The remainder of this paper is organized as follows:
In Section ~\ref{Experiment} we summarize the experimental 
results for YMnO$_3$ from the literature.
In Section ~\ref{theorysection} we briefly describe our plane wave 
pseudopotential implementation of density functional theory
and outline the technical details of the calculations
used in this work.
In Section~\ref{Results} our results for
cubic and hexagonal YMnO$_3$ are presented, followed by a comparison
of the properties of
cubic YMnO$_3$ with those of LaMnO$_3$.
Our results are summarized in Section~\ref{conclusions}.

\section{Summary of experimental results}
\label{Experiment}

The literature on the hexagonal manganites is much more
limited than that on the cubic perovskite
manganites, and modern work
focuses
largely on yttrium manganite, YMnO$_3$. Although yttrium is not
a rare earth, it forms a stable trivalent cation with a similar
ionic radius to those of the smaller rare earth ions.
It is favored over the smaller rare earth ions in research studies for
two reasons. First,
study of magnetic ordering on the Mn$^{3+}$ ions is more
straightforward in YMnO$_3$ than in the rare earth manganites
since there is no perturbation from the rare earth $f$-electron magnetic
moment. In addition, YMnO$_3$ forms both the hexagonal and 
orthorhombic phases on solution synthesis at ambient pressure\cite{Yakel}
making it particularly appealing for use in a comparative study.

Early (1960s) work on YMnO$_3$\cite{Yakel} established the hexagonal
phase to be ferroelectric, with the hexagonal perovskite structure and
the $P6_3cm$ space group.
The hexagonal perovskite structure consists of $ABCACB... $ stacking of close
packed O layers, with the Mn ions occupying 
5-fold
coordinated sites, and the rare earth atoms in 7-fold coordinated
interstices. The ideal high temperature paraelectric structure
consists of two formula units per unit cell and is shown in Figure
~\ref{hexperovskite}.
Since this
structure does not occur in the perovskite ferrites it has been
suggested that the ability of Mn$^{3+}$ to form 5-fold trigonal
bipyramids through $dsp^3$ hybrid bonding might be a requirement\cite{Yakel}.
In addition (not shown in the Figure) at low temperature the MnO$_5$ 
bipyramids are 
slightly rotated around the axis passing through the Mn and parallel to 
one of the triangular base sides.
Early reports of a weak parasitic ferromagnetism\cite{Bertaut1}
were soon shown to be the result of Mn$_3$O$_4$ impurity 
contaminating the powdered phase\cite{Bertaut2}, and weak
ferromagnetism was shown to be absent in single crystal 
samples\cite{Bertaut2}.

Modern studies of hexagonal YMnO$_3$ have established that the ferroelectricity
occurs along the $c$ axis (i.e.[0001])\cite{fiyi,klk}, with 
a spontaneous ${\bf P}$ $\sim$ 5.5 $\mu$C/cm$^2$
In addition a strong
coupling between the ferroelectric and
magnetic ordering has been revealed\cite{Huang}.
Although the ferroelectric Curie temperature (914K) is quite
different from the N\'{e}el temperature (80K), anomalies in the
dielectric constant and loss tangent near the N\'{e}el temperature
are indicative of coupling between the ferroelectric and antiferromagnetic
orders. The coupling has been attributed to changes in the phonon
spectrum associated with the antiferromagnetic transition\cite{Huang}.
Raman and infra-red spectroscopy of the high temperature paraelectric
and low temperature ferroelectric phases show only weak bands in the
ferroelectric phase due to the non-centrosymmetricity. This indicates
that the structural differences between 
the ferroelectric and paraelectric phases are very small. 
Non-linear optical spectroscopy\cite{Frohlich} shows two types
of optical second harmonic spectra of the Mn$^{3+}$ ions in 
hexagonal YMnO$_3$, one from the non-centrosymmetric ferroelectric
ordering of charges, and the other caused by the centrosymmetric
antiferromagnetic ordering of spins. Partial overlapping between
the electronic transitions results in a nonlinear optical
polarization which depends on two order parameters.
A generalized Ginzburg-Landau formalism has been developed which
shows that the second harmonic generation susceptibility is
directly proportional to the bilinear combination of both order
parameters\cite{Sa}.
Finally, epitaxial thin films of hexagonal YMnO$_3$ have been
grown on (111) MgO and (0001) ZnO:Al/(0001)\cite{fiyi}, 
and explored for use as non-volatile memory devices.
YMnO$_3$ presents some technological advantages over many common 
ferroelectric perovskites, including lower dielectric constant 
($\sim$ 20 at room temperature) and non-volatile constituent elements. 

In the 1970s, work began on the orthorhombic phase of YMnO$_3$. 
Susceptibility
measurements established that orthorhombic YMnO$_3$ is an
antiferromagnet with a N\'{e}el temperature of 42K, although
significant deviations from Curie-Weiss behavior were seen 
well above the manganese ordering temperature\cite{Wood}. A more detailed
study\cite{Quezel} concluded that the Mn$^{3+}$ ordering is helical with a
propagation wavevector {\bf k} = [0,0.0786,0] and a helical
angle from plane to plane of 14$^{\circ}$. Within each plane
the ordering is antiferromagnetic.

There has also been some modern work on the orthorhombic phase
of YMnO$_3$. Attempts have been made to find a synthesis which
will produce a larger proportion of the metastable orthorhombic
phase\cite{Brinks}. In addition the Raman spectrum of orthorhombic
YMnO$_3$ has been
measured and the Raman active phonon modes determined and compared with those
of LaMnO$_3$\cite{Iliev}.

\section{Computational Technique}
\label{theorysection}

The calculations described in this work were performed using
plane wave pseudopotential (PWPP) implementations\cite{Yin} of 
density functional 
theory\cite{HKKS} within the local spin density approximation (LSDA). 
The accuracy and efficiency of {\it ab initio} pseudopotential
calculations (compared with all-electron calculations) is now
well established for spin-polarized magnetic systems,
and the PWPP method has been applied to a wide range of magnetic
materials\cite{PWPPegs}.

In this work we use both
the optimized pseudopotentials developed by Rappe et al.\cite{Rappe}
and the Vanderbilt ultra-soft pseudopotentials\cite{Vanderbilt},
to reduce the energy cut-off of the plane wave
expansion to around 60 and 35 Ry respectively. 
For La and Y, we constructed scalar-relativistic 
pseudopotentials,  and 
for Mn and O non-relativistic pseudopotentials were used
with the partial
non-linear core correction scheme of Louie et al.\cite{NLCCs} applied
to the Mn ions. 
The pseudopotentials were
tested for transferability by comparing with all-electron calculations
for a range of typical atomic and ionic configurations. The 
pseudo-eigenvalues and total energies were found to be
equivalent to the all-electron
values to within a few meVs.
All pseudopotentials were converted into the usual
Kleinman-Bylander form\cite{Kleinman_Bylander}, and
the absence of ghost states was confirmed using the ghost theorem
of Gonze, K{\" a}ckell and Scheffler\cite{Gonze}.
The pseudopotential construction and the total energy calculations
used the Perdew-Zunger parameterization\cite{Perdew_Zunger} of the 
Ceperley-Alder exchange correlation potential\cite{Ceperley_Alder}
with the von Barth-Hedin interpolation formula\cite{vonBarth_Hedin}.

The total energies and band structures were calculated using 
the CASTEP 2.1\cite{Payne,Castep} and PWSCF ultra-soft
pseudopotential programs.
For cubic perovskites we used a plane wave cut 
off of 60 Ry, which corresponds to around 3500 plane waves in a 
cubic unit cell with lattice constant of around 4 \AA\ . 
A 6x6x6 Monkhorst-Pack\cite{Monkhorst_Pack} grid was used for all
calculations for cubic systems. This led to 10 k-points in the 
irreducible Brillouin Zone
for the high symmetry structures, and a correspondingly
higher number for the structures with lower symmetry. 
In calculations for the hexagonal structure, a 5x5x4 Monkhorst-Pack
grid was used, resulting in 18 k-points in the irreducible Brillouin Zone
for the high symmetry structures.
A variable Gaussian broadening between 1 eV and 0.002 eV was applied 
to the k-point sampling to speed convergence for metallic systems.
For density of states (DOS) calculations we 
calculated first-principles eigenvalues of a large ($\approx$ 100 
in the irreducible Brillouin zone) 
k-point set, then interpolated using the 
interpolation scheme of Monkhorst and Pack\cite{Monkhorst_Pack}. 
We then applied the 
Gilat-Raubenheimer method\cite{Gilat-Raubenheimer} to integrate
over this fine 
mesh. Finally, for the band structure plots, and for use in the
tight-binding analysis, symmetry labels along high-symmetry lines 
were assigned using projection operators for the corresponding
irreducible representations.

\section{Results}
\label{Results}

\subsection{Calculated Electronic Properties of cubic and hexagonal
YMnO$_3$}
In this section we compare the electronic properties of YMnO$_3$ in
its metastable cubic perovskite structure with those of the
stable hexagonal phase.
We begin by calculating the electronic structure
for the ideal high symmetry structures, without including magnetic effects,
then lower the magnetic symmetry to the hypothetical ferromagnetic  and
observed 
antiferromagnetic phases. Finally we introduce structural distortions
to simulate the effects of the onset of ferroelectricity.
This ability to isolate structural and magnetic
distortions is unique to computational studies,
and allows identification of the essential microscopic
interactions which cause the observed macroscopic behavior.

\subsubsection{High Symmetry Paramagnetic Structures}

First we present the calculated electronic structures for
YMnO$_3$ in the ideal cubic and hexagonal structures, 
without spin polarization (we call this the paramagnetic (PM)
phase). 

\paragraph{Cubic}
Figure~\ref{cubic_DOS} shows the calculated density of states
(DOS) for cubic paramagnetic YMnO$_3$. 
We chose a lattice constant
of 3.84 \AA\ which corresponds to the average of the lattice 
constants in the low temperature orthorhombic unit cell.
The plotted energy
range is from -8 eV to 4 eV, and the lower lying semi-core
states have been omitted for clarity.
The Fermi level is set to zero. 
The broad series of bands between 
approximately $-2$ and $-7$ eV arises primarily
from the oxygen $2p$ orbitals. Above the oxygen $2p$ bands, and
separated from them by an energy gap, are the Mn $3d$ bands. The
Mn $3d$ bands are divided into two sub-bands - the lower energy
$t_{2g}$ bands, and the higher energy $e_g$ bands - as a result
of the crystal field splitting by the octahedral oxygen anions.
The Fermi level lies near the top of the Mn $3d$ $t_{2g}$ bands
and is in a region of high density of states.
The large DOS at the Fermi level confirms that the cubic PM structure
is unstable, and that a lower energy structure could be achieved by
allowing spin-polarization and/or structural distortion.

\paragraph{Hexagonal}
The hexagonal paraelectric structure has three lattice parameters;
the usual $a$ and $c$ 
parameters plus an internal $u$ that gives the distance (in units of $c$) 
between oxygen and yttrium layers, 
and is not fixed by the symmetry. 
In {\it ideal} hexagonal structures (without tilting of the
oxygen octahedra) the atomic positions in crystal coordinates are:
Y at (0,0,0) and (0,0,$\frac{1}{2}$), 
Mn at ($\frac{1}{3}$,$\frac{2}{3}$,$\frac{1}{4}$)
and ($\frac{2}{3}$,$\frac{1}{3}$,$\frac{3}{4}$)
O at (0,0,$\frac{1}{4}$), (0,0,$\frac{3}{4}$),
($\frac{1}{3}$,$\frac{2}{3}$,$u$),($\frac{2}{3}$,$\frac{1}{3}$,$1-u$),
($\frac{2}{3}$,$\frac{1}{3}$,$\frac{1}{2}+u$) and
($\frac{1}{3}$,$\frac{2}{3}$,$\frac{1}{2}-u$), with $u$ not fixed
by the symmetry.
By energy minimization we obtain $a$=3.518 \AA, $c$=11.29 \AA, 
and $u$=0.084, in very good agreement with the experimental values 
$a$=3.539 \AA, $c$=11.3(4) \AA, 
and $u$=0.084. This suggests that the oxygen tilting which we are
neglecting in these calculations indeed has little 
effect on the structural properties.
Note however that in the ideal, non-magnetic, high symmetry structures, 
we find that the hexagonal
phase is not energetically stabilized over the cubic phase. This
indicates that the low temperature structural distortions and magnetic 
polarizations which we investigate in the next 
section are responsible for the stability 
of the hexagonal phase.

In Figure~\ref{fig5} we show the calculated
orbital-resolved DOSs for the non-magnetic
hexagonal phase at the LSDA minimum energy lattice parameters.
For clarity only the DOS on the {\it transverse} oxygens is shown 
since these are most strongly involved in chemical bonding with
the Mn ion and will be most important in our later analysis of
ferroelectricity.
As in the cubic case, the non-magnetic hexagonal 
phase shows a large DOS at E$_F$, mostly due to the Mn 
$d$ states located 
in a narrow region around E$_F$. In sharp contrast with the cubic phase,
however, the hexagonal crystal field splits the 
$d$ states into three groups. The $d_{xy}$ and 
$d_{x^2-y^2}$ orbitals lie on the (0001) plane and 
are nearly 
degenerate (only one of them is shown in the Figure). These
orbitals point towards the nearby in-plane oxygens (O$_P$), 
and so are able to participate in covalent bonding, therefore they
form a rather broad band. 
The $d_{xz}$ and $d_{yz}$ orbitals are also 
almost degenerate. Their DOSs overlap with those of $d_{xy}$ and $d_{x^2-y^2}$, 
but are localized in a narrower 
energy region, since they do not point directly towards oxygens and 
therefore their bands are not broadened by hybridization.
Each orbital of the two doublets is roughly half occupied. 
Finally, $d_{z^2}$ (shaded gray in the Figure for clarity)
is the highest in energy and the least occupied.
It points towards the transverse oxygens (O$_T$) which are the 
closest ligands to the Mn ions and therefore cause the strongest crystal
field destabilization. 
(The distance between the Mn ion and the transverse oxygens along the
$c$ axis is 1.875 \AA\, and that between 
the Mn ion and the in-plane oxygens is 2.03 \AA\ .)
However it is important to note that a 
certain amount of $d_z{^2}$ charge is hybridized with the O$_T$ $p_z$ states 
located $\sim$ 5 eV below E$_F$. 

In the next section
we will show that these important differences in the splitting of the Mn $d$
bands resulting from the different Mn coordination in the two structures,
which are apparent already in the high symmetry paramagnetic phases,
have a profound effect on their subsequent electronic, structural and
magnetic properties in the realistic spin-polarized structures.

\subsubsection{High symmetry structural phases with magnetic polarization}

Next we calculate the properties of both cubic and hexagonal phases, still 
with the ideal structural high symmetry, but this time in the 
experimentally observed antiferromagnetic ground state.
In both cases, allowing spin polarization lowers the total
energy, consistent with the Stoner model, and introduces magnetic moments
onto the Mn ions. 

\paragraph{Cubic}

In Figure~\ref{AFM.cub.dos} we show the orbital resolved densities
of states on single Mn and O ions,
in ideal cubic A-type antiferromagnetic YMnO$_3$ at the same
lattice constant used for the paramagnetic calculations.
The A-type AFM phase
consists of ferromagnetically aligned (100) planes of Mn ions coupled
antiferromagnetically to each other.
Due to the exact AFM symmetry, each atom has a corresponding 
atom in the cell with anti-aligned spin, so the up and down 
components of DOS are exchanged.
The upper plot shows the up-spin and down-spin
Mn $3d$ densities of states above and below the x-axis respectively,
and the lower plot shows the up-spin oxygen density of
states.  The down-spin oxygen density of states is not shown since
it is very similar to the up-spin.
The oxygen density of states in this region derives almost entirely
from the oxygen $2p$ orbitals; there is negligible contribution from 
the O $2s$ orbitals in the region shown.
The densities of states of the $d_{xy}$, $d_{xz}$ and
$d_{yz}$ (not shown) orbitals are very similar. In the absence of
the antiferromagnetic ordering these would be identical by symmetry
and would form the $t_{2g}$ manifold. Similarly the $d_{z^2}$ and
$d_{x^2-y^2}$ bands, from the $e_g$ manifold, are alike.
The $t_{2g}$ band is rather narrow for both spin directions, and
clearly shows an exchange splitting of around 2 eV. This splitting 
causes the majority $t_{2g}$ band to be fully occupied while the 
minority $t_{2g}$ band contains only a small amount of electron density. 
The $e_g$ band is much broader since the $d_{z^2}$ and $d_{x^2-y^2}$
orbitals are more able to hybridize with their oxygen neighbors. It 
is roughly half-occupied in the majority case, but completely unoccupied in 
the minority case. 
The calculated Mn magnetic moment of around 3 $\mu_B$, significantly
reduced from the Hund's rule value of 4 $\mu_B$ expected for a $d^4$ ion,
reflects the fact that the exchange splitting is not large enough to
produce full spin polarization.
It is interesting to note that the region of overlapping energy between
the Mn $3d$ and oxygen $2p$ states is in fact rather small, with only
a small amount of Mn density occurring in the oxygen rich region 
between around -3 and -9 eV, and vice versa.
The similarity between the up- and down- spin oxygen densities of states
is a a consequence of the small Mn $3d$ - O $2p$ overlap.

For comparison, the orbital resolved densities of states in ideal cubic 
{\it ferromagnetic} YMnO$_3$ are shown in Figure~\ref{FMcubic_DOS}. The 
upper plot shows the up-spin and down-spin Mn $3d$ densities of states 
above and below the x-axis respectively, and the lower plot shows the 
up- and down-spin oxygen densities of states. In this case we have
plotted only one of the $t_{2g}$ and one of the $e_g$ bands for
clarity. The only point that we want to make here is that the
densities of states are very similar to those of the antiferromagnetic
structure. Therefore as long as spin polarization is allowed, the
electronic structure is not very sensitive to the {\it relative} polarization
of neighboring Mn ions.

\paragraph{Hexagonal}

The paraelectric hexagonal structure contains two Mn ions per unit cell, 
allowing studies of FM and AFM spin orientations without employing
supercells. We find that the AFM phase is slightly lower in energy than 
the FM phase, as is the case in experiment. We focus on
the AFM phase in what follows although, as we saw already in the 
cubic case, 
the spin ordering is not essential for our arguments. The 
properties that we consider here
are determined by the local (i.e. internal to the oxygen cage) spin 
polarization rather than by 
long-range magnetic ordering.

In contrast to the non-spin-polarized phase, in which the 
$d_{xz}$, $d_{yz}$ and $d_{xy}$, $d_{x^2-y^2}$ doublets were
roughly half occupied, here
the Stoner exchange produces a 
$\sim$ 2.5 eV energy splitting that causes these four orbitals to become 
almost completely spin 
polarized (Figure~\ref{hex.afm.dos}). Their total up and down charges 
are indeed 
3.93 and 0.47 electrons,
respectively, while the occupation numbers of $d_{z^2}^{\uparrow}$ and 
$d_{z^2}^{\downarrow}$ 
orbitals are 0.53 and 0.30 electrons. The resulting magnetic moment on 
Mn (3.7 $\mu_B$) is less than the 4 $\mu_B$ Hund's rule value
because of the 
hybridization of $d_z^2$ with
O$_T$ $p_z$ states, and of $d_{xy}$ and $d_{x^2-y^2}$ with O$_P$ $p_x$ and 
$p_y$ states. However it is larger than the value calculated for the
cubic case because of the larger exchange splitting occurring here.
A non-negligible magnetic moment, M=0.15 $\mu_B$, is also present on O$_P$, 
driven by 
exchange with the $d_{xy}$ and $d_{x^2-y^2}$ orbitals, whereas the 
magnetization of O$_T$ is 
almost vanishing, due to the AFM symmetry (for $u$=0 it would be exactly 
zero).  
 
The large spin splitting does not manage to open a gap. Indeed there is a 
tiny DOS at E$_F$,
to which $d_{xy}$ and $d_{x^2-y^2}$ orbitals from Mn, and $p_x$ and $p_y$ 
orbitals from O$_P$ 
contribute. This means that the bands crossing E$_F$ come completely from 
orbitals oriented in the $x-y$ plane
(the tilting of the oxygen cages in the ferroelectric structure is probably 
sufficient to open an  energy gap).  
The mainly planar distribution of the charge density is also evident from 
the band energies
shown in Figure~\ref{fig7}, calculated within the $k_z$=0 plane of the 
Brillouin zone (left panel), 
and along [0001] (right panel). Only one band crosses E$_F$ at $k_z$=0, 
and its orbital character is a 
mix of $d_{xy}^{\uparrow}$ and $d_{x^2-y^2}^{\uparrow}$ states from Mn, 
and $p_x^{\uparrow}$ and 
$p_y^{\uparrow}$ states from O$_P$.  
The band energies are extremely flat along [0001], as is typical for a 
strongly layered compound. The system can
be described as a poor metal in the $x-y$ plane, and as an insulator 
perpendicular
to it. If we discard the first unoccupied bands 
coming from in-plane orbitals, we obtain a one-dimensional `pseudogap' 
of $\sim$ 2 eV between occupied O$_T$ p 
and empty Mn $d_{z^2}$ states, similar to the energy gap present between
occupied oxygen states and empty transition metal $d$ states in classic
ferroelectric perovskite oxides such as BaTiO$_3$. 

\subsubsection{Effects of ferroelectric distortion on the electronic properties of cubic and hexagonal structures}

Finally we investigate the sensitivity of the charge distribution to
atomic displacements by moving the Mn and in-plane oxygen atoms 
in opposite directions along the $c$ and $z$ axes in hexagonal and 
cubic phases respectively. 

The results for the hexagonal structure are shown in Figure~\ref{fig10}.
Note that since the electric polarization occurs 
along the $c$ axis, and the 
bipyramidal tilting does not alter significantly the features of this compound,
the paraelectric 
structure without the addition tilting is a sufficient starting
point for the purpose of studying the nature
of the ferroelectric displacement. 
For clarity, only the Mn orbitals most involved in the 
change of hybridization (i.e. $d_{xz}$ and $d_{z^2}$) are shown. The change 
of DOS upon distortion 
is very evident. The Mn $d$ DOS is shifted down by $\sim$ 2 eV and is 
strongly hybridized 
with the O$_T$ $p$ states that are localized in a $\sim$1 eV wide energy 
region.  The atomic displacements increase the overlap of 
$d_{z^2}$ and O$_T$ $p_z$ orbitals, inducing a $pd_{\sigma}$ hybridization, 
and the overlap of $d_{xz}$ and 
$d_{yz}$ with O$_T$ $p_x$ and $p_y$ orbitals, increasing the $pd_{\pi}$ 
hybridization.  Note that very similar changes in hybridization with 
ferroelectric displacement are seen
in the prototypical ferroelectric BaTiO$_3$\cite{Alessio}. However in 
BaTiO$_3$ all the Ti $3d$ orbitals are formally empty, and so displacement 
in {\it any} direction causes an energy lowering rehybridization as the oxygen 
ions transfer charge into the empty $d$ orbitals. By contrast, the 
ferroelectricity in YMnO$_3$ is strongly uniaxial because only the 
$d_{z^2}$ orbital is empty. Since the majority spin $d_{xz}$ and
$d_{yz}$ orbitals are almost completely filled, their changes of hybridizations
with oxygen on displacement, although significant, do not contribute to a change
in electric polarization.

The markedly different results which we obtain for the cubic structure
are shown in Figure~\ref{cub.dis.dos}.
Again only those orbitals most affected by the distortion (Mn $d_{z^2}$ and
$d_{xz}$ and O$_T$) are shown. The changes in DOS on distortion are
much less pronounced and quite different in character from those which
we found for the hexagonal phase.
The most obvious change is a destabilization of the $d_{xz}$ band which
results from the larger crystal field repulsion as the Mn is moved
closer to the transverse oxygens along the $z$ axis. Since the $d_{xz}$ 
orbitals are part of the filled $t_{2g}$ manifold they are unable
to lower their energy by rehybridization with the oxygen $2p$ states,
therefore the only contribution to their energetics is the 
unfavorable crystal field repulsion. Note that a typical Jahn-Teller
distortion, in which the oxygens along the $z$ axis move {\it away}
from the cation, would have the opposite effect.
In addition there is a slight broadening and shift down in energy
of the $d_{z^2}$ band, which, since it was originally partly
unoccupied, is now better able to accept charge density from the
oxygen ligand that it has moved closer to.
There is an almost rigid shift of the oxygen $2p$ bands to lower
energy relative to the Fermi energy. This in fact a
consequence of the shift of the Fermi level to higher energy due
to the destabilization of the Mn $3d$ bands.
The fact that the form of the
oxygen bands does not change significantly is an indication that 
rehybridization is minimal. 

\subsection{Comparison of the electronic properties of cubic YMnO$_3$
and LaMnO$_3$}

Finally we briefly compare our calculated band structures of YMnO$_3$ in
its metastable cubic perovskite structure, with those of the
prototypical cubic manganite, LaMnO$_3$. The purpose of this
section is to determine whether there are any important differences
between the {\it electronic} properties of cubic YMnO$_3$ and
LaMnO$_3$ which might explain why the cubic phase is stable
in LaMnO$_3$ but metastable in YMnO$_3$.
(Note that the cubic phase does not support ferroelectricity for
either compound, therefore since
LaMnO$_3$ does not form in the hexagonal structure it is {\it
never} ferroelectric.) In fact we will find that the 
band structures of cubic YMnO$_3$ and  LaMnO$_3$ are very
similar, and therefore conclude that LaMnO$_3$ only forms
in the cubic phase, whereas YMnO$_3$ can be either cubic
or hexagonal (and therefore ferroelectric) because the
yttrium ion is smaller than the lanthanum ion.

First we report in Table~\ref{table1} the relative energies of 
the paramagnetic (PM), ferromagnetic (FM) and A-type 
(AFM) phases of cubic YMnO$_3$ and LaMnO$_3$. 
The relative energies of FM and AFM phases of LaMnO$_3$ are
taken from Ref. \cite{Pickett_Singh} and those for YMnO$_3$ were 
calculated in this work.
The most important feature for our purposes is that the relative 
orderings of the states is the same in YMnO$_3$ as in LaMnO$_3$.
It is also striking
that in both cases the ferromagnetically ordered state is the most stable
phase. This has been pointed out previously for LaMnO$_3$
\cite{Pickett_Singh}, and the relationship between the observed
Jahn-Teller distortion and the resulting A-type antiferromagnetic
ground state is well understood \cite{Goodenough}.
Also, in both cases the calculated magnetic moment is
about 3.0 $\mu_{\rm B}$ per Mn for the cubic FM and AFM structures,
lower
than the spin-only value of 4 $\mu_B$ for the Mn$^{3+}$ ion as
discussed earlier.
The similarity in magnetic moment indicates similar
amounts of spin-polarization and hybridization in both cases.

Figure~\ref{PM_BS} shows the band structures of cubic PM YMnO$_3$
and LaMnO$_3$ plotted
along the high symmetry axes of the simple cubic Brillouin Zone. 
The Fermi level is set to 0 eV in both cases.
The broad O $2p$ bands between around -2 and -7 eV can be seen
clearly in both materials, with the Mn $3d$ bands above them,
separated by an energy gap. 
In LaMnO$_3$, the Mn $3d$ bands are largely separated in energy
at each $k$-point from the next highest energy bands, which are
the La 5$d$ bands. The highlighted lines in the LaMnO$_3$ band
structure plots accentuate the upper and lower Mn $3d$ bands, which
have a similar form to each other and have been seen in other
cubic perovskite manganites
(see Refs. \cite{HillRabe} and \cite{Pickett_Singh}.) This indicates a
`universality' in the
manganite structure throughout the cubic perovskite manganite series,
which is independent of the identity of the large cation. 
The same structure for the upper and lower Mn $3d$ bands can also 
be identified in the YMnO$_3$ band structure,
however the
Mn $3d$ bands are slightly broader than those in LaMnO$_3$ because
the smaller lattice constant results in greater Mn $3d$ - O $2p$
overlap. The smaller lattice constant, and correspondingly larger
crystal field repulsion, is also responsible for the larger $t_{2g}$ -
$e_g$ splitting which can be clearly seen at the $\Gamma$
point.
In addition the Y $4d$ bands are lower in energy
than the La $5d$ bands. These factors result in the lower part of
the Y $4d$ bands intersecting the upper part of the Mn $3d$
bands in some regions of the Brillouin zone, making the universal structure
less well-defined.

To quantify any differences between cubic PM YMnO$_3$ and LaMnO$_3$,
we performed tight-binding analyses of the $\Gamma$ to X regions of 
the respective band structures. Tight-binding parameters were obtained by 
non-linear-least-squares fitting\cite{Mattheiss} to the calculated 
{\it ab initio} energies at the high symmetry $\Gamma$ and X points, and 
at 19 points along the $\Delta$ axis.
The tight-binding parameters thus obtained 
are given in Table~\ref{TB_params1}.  
In both cases a good (root mean square (RMS) deviation
from the {\it ab initio} energies of 0.20)
fit was obtained by including only oxygen $2s$ and $2p$ and Mn $3d$ 
orbitals in the basis set, since 
we find no significant
covalent bonding between the A cation and either the oxygen
$2p$ or the Mn $3d$ orbitals.
This is
consistent with an early proposal by Goodenough\cite{Goodenough}
that the magnetic properties of the rare earth manganites are determined 
by the Mn $3d$ - O $2p$ hybridization only. 
The parameters for YMnO$_3$ are
similar to those for LaMnO$_3$, except for an increase of around 10 \%
in the Mn $3d$ - O nearest neighbor overlaps. This is a consequence
of the smaller lattice constant in YMnO$_3$, and results in the
slightly broader Mn $3d$ bands noted above.

To test the effect of size on the electronic properties, we also calculated the
properties of cubic paramagnetic YMnO$_3$ at the experimental
LaMnO$_3$ lattice constant of 3.95 \AA\ . The resulting band structure 
of this hypothetical structure (not shown) is
qualitatively similar to 
that of YMnO$_3$ at
the true experimental lattice constant, except for a slight narrowing of the
bandwidths as a result of the reduced overlap between the atomic wavefunctions,
and a reduced $t_{2g}$ - $e_g$ splitting as a result of the lower crystal
field repulsion.

We find that 
introduction of spin polarization causes the same kinds of changes in
both compounds, therefore the conclusions drawn from analysis of the
paramagnetic band structures continue to be valid.
Figure~\ref{FM_BS} shows the up and down-spin band structures
for YMnO$_3$ and LaMnO$_3$ along the high symmetry axes of
the simple cubic Brillouin Zone. 
(We chose the ferromagnetic (FM) phases
for simplicity in interpreting the band structures; 
as discussed above, we are interested
in the effects of spin polarization rather than the nature of the
ordering between magnetic moments.) 

First we examine the up- and down-spin LaMnO$_3$ band structures,
and compare with the PM LaMnO$_3$ band structures to determine the 
changes which spin polarization causes in a `conventional' manganite. 
The states which correspond to non-magnetic atoms are unchanged from
the paramagnetic state, and are identical for up- and down-spin
electrons. For example,
the dispersion of the lowest O $2p$ band is very similar for up- and
down-spin, and for the PM phase. Also the La $5d$ bands, which
were above the Mn $3d$ bands in the PM state are unchanged in 
form and energy.
The characteristic perovskite manganite Mn $3d$ pattern
which we remarked on earlier persists in the FM phase, appearing
around 2 eV higher for the down-spin electrons than for the 
up-spin because of the exchange splitting. The up-spin Mn $3d$ and
O $2p$ bands are strongly hybridized and there is no gap between them.
However, the down-spin Mn $3d$ are split off from the O $2p$ bands
by a larger gap than in the PM case. As a result, the Mn $3d$ bands 
occupy the same energy region as the unoccupied La $5d$ bands. 
The Fermi level cuts through the
very bottom of the down-spin Mn $3d$ bands, and the
conduction band is occupied almost entirely by up-spin electrons,
indicating half-metallic behavior.

Finally we compare the up- and down-spin YMnO$_3$ band structures.
The first notable feature is that the non-magnetic Y $4d$ bands 
and O $2p$ bands are very similar for {\it both} up- and down- spin. 
This is because 
the exchange interaction in this hypothetical ferromagnetic 
YMnO$_3$ structure does not remove the gap between
the majority Mn $d$ bands and
the O $2p$ band. 
(By contrast, in LaMnO$_3$ the higher O $2p$ bands overlap
with the majority Mn $3d$ bands and therefore show a polarization
dependence). 
This is again the result of the
smaller lattice constant in YMnO$_3$, which destabilizes the
unfavorable FM interactions compared with those in LaMnO$_3$,
where the Mn ions are further apart.
Both up- and down-spin
Mn $3d$ bands show the characteristic perovskite manganite
features (most noticeable in this case along the M-R-X directions) which we 
noted earlier in our discussions of the PM band structures.

The strong similarities between the calculated properties of cubic
YMnO$_3$ and LaMnO$_3$ that we have presented in this section
confirm that the observed differences between 
the properties of hexagonal and cubic manganites
are merely the result of the different
sizes of the large cations in the two materials stabilizing different
structures, and not due to any additional
covalent bonding effects.            

\section{Conclusions}
\label{conclusions}

In summary, the formal similarity of $d^4$ Mn$^{3+}$ in hexagonal and
cubic YMnO$_3$ does not prevent the two phases from showing quite
different properties. In particular, the difference between hexagonal
and cubic field symmetry induces a very different charge distribution.
In cubic perovskites, the Mn$^{3+}$ ion has partially occupied $e_g$
states. This creates an instability relieved by Jahn-Teller distortion and
the ordering of Mn orbitals on the (001) plane. Both Jahn-Teller and
orbital ordering tend to favor structural configurations that keep Mn and O
aligned on the same plane, and therefore do not result in ferroelectricity.
In the hexagonal perovskites, the crystal
field produces an ordering of $d$ states that leaves the $d_{z^2}$
orbital mostly unoccupied, and thus able to hybridize with the $p_z$
orbital of the transverse oxygen. This charge environment favorable to
electric polarization along the $c$ axis is only realized in condition
of spin-polarization, not for the non-magnetic hexagonal phase.
Thus, in hexagonal YMnO$_3$ the spin-polarization is actually
the factor which enables the ferroelectricity.
Finally, we emphasize that the transition from the AFM to the paramagnetic
phase above T$_N$=80 K
(i.e. well below the critical temperature of the ferroelectric phase), does
not invalidate our picture,
since it is the local spin-polarization of Mn$^{3+}$ that matters,
independently of the actual presence (or absence) of spin ordering.
Note also that there are no fundamental differences between 
the electronic structures of metastable cubic YMnO$_3$ and the prototypical
stable cubic manganite, LaMnO$_3$. Therefore the metastability of the
cubic perovskite phase in YMnO$_3$ 
is simply the result of the smaller size of the A cation  which is unable
to stabilize the cubic phase. 

\section{Acknowledgments}

The authors acknowledge many useful discussions with 
Profs. Warren Pickett and  Karin Rabe.
We thank the authors of CASTEP for providing us with their
software package, and Umesh Waghmare for implementing the
extension to metallic systems.
Funding for this work was provided by the National Science Foundation,
grant numbers  DMR-9973076 and DMR 9973859.
Calculations have been carried out on the IBM SP2 machine of the MHPCC 
Supercomputing Center in Maui, HI.

\clearpage

\pagebreak
 
\noindent
{\bf Figure Captions}
 
\begin{figure}
\caption{The ideal
cubic perovskite structure. The small B cation (in black) is at
the center of an octahedron of oxygen anions (in gray). The large A cations 
(white) occupy the unit cell corners.}
\label{perovskite}
\end{figure}

\begin{figure}
\caption{The ideal
hexagonal perovskite structure. The small B cation (in black) is 
five-fold coordinated by oxygen anions (in gray). The dashed lines indicate
the oxygen pentahedra. The large A cations (white)
occupy seven-fold coordinated sites.}
\label{hexperovskite}
\end{figure}

\begin{figure}
\caption{Calculated density of states for cubic paramagnetic
YMnO$_3$.}
\label{cubic_DOS}
\end{figure}

\begin{figure}
\caption{Orbital resolved densities of states
for Mn and transverse oxygen in the paraelectric 
non-magnetic phase of hexagonal
YMnO$_3$. Only one DOS for each of the nearly degenerate pairs of orbitals 
($d_{xy}$,$d_{x^2-y^2}$) and
($d_{xz}$, $d_{yz}$) is shown.}
\label{fig5}
\end{figure}

\begin{figure}
\caption{Orbital resolved densities of states on single Mn and O atoms
in cubic antiferromagnetic YMnO$_3$.} 
\label{AFM.cub.dos}
\end{figure}

\begin{figure}
\caption{Orbital resolved density of states in cubic ferromagnetic 
YMnO$_3$. }
\label{FMcubic_DOS}
\end{figure}

\begin{figure}
\caption{Orbital resolved DOS of a single Mn ion (top panel), in-plane
oxygen O$_P$ (middle), and transverse oxygen O$_T$ (bottom)
in the paraelectric AFM phase of hexagonal
YMnO$_3$. p$^u$ and p$^d$ indicate up and down components of O p DOS, 
respectively. }
\label{hex.afm.dos}
\end{figure}

\begin{figure}
\caption{Band energies of paraelectric AFM hexagonal YMnO$_3$. Q=$\pi$/a 
(0,2/3$^{1/2}$,0), P=$\pi$/a 
(1,1/3$^{1/2}$,0)=$\Gamma$.}
\label{fig7}
\end{figure}

\begin{figure}
\caption{Orbital-resolved DOS for single Mn and O in hexagonal AFM YMnO$_3$. 
The upper panel is for the paraelectric 
structure, the lower for the structure with Mn and O displaced along the $c$ 
axis, in the fashion of a ferroelectric 
distortion.}
\label{fig10}
\end{figure}

\begin{figure}
\caption{Orbital-resolved DOS for single Mn and O in cubic A-type
AFM YMnO$_3$.  The upper panel is for the paraelectric 
structure, the lower for the structure with Mn and O displaced along the z 
axis, in the fashion of a ferroelectric 
distortion.}
\label{cub.dis.dos}
\end{figure}

\begin{figure}
\caption{Calculated band structures for cubic paramagnetic
LaMnO$_3$ and YMnO$_3$ along the high symmetry axes of the
Brillouin Zone.
The highlighted lines in the band
structure plots accentuate the upper and lower Mn $3d$ bands, which
have a similar form to each other and to those of other perovskite 
manganites.}
\label{PM_BS}
\end{figure}

\begin{figure}
\caption{Up- and down-spin band structures for cubic LaMnO$_3$
and YMnO$_3$ along the high symmetry axes of the
Brillouin Zone.}
\label{FM_BS}
\end{figure}

\clearpage

\begin{table}
\begin{center}
\begin{tabular}{|| l | r | r ||}
                           &  LaMnO$_3$ & YMnO$_3$   \\
$E_{O2s}$                  & -17.836376 & -17.9649   \\
$E_{O2p}$                  &  -4.514022 &   -4.7382  \\
$E_{Mn3d}$                 &  -1.191285 &   -1.4097\\
$V_{O2s-O2s}$              &  -0.243683 &   -0.2461\\
$V_{(O2p-O2p)\sigma}$      &   0.620568 &   -0.7167 \\
$V_{(O2p-O2p)\pi}$         &  -0.063466 &   -0.1069\\
$[V_{(O2p-O2p)\sigma}]_2$  &   0.182635 &    0.1688\\
$[V_{(O2p-O2p)\pi}]_2$     &   0.082951 &    0.0620\\
$V_{O2s-Mn3d}$             &  -1.735814 &   -1.9711\\
$V_{(O2p-Mn3d)\sigma}$     &  -1.838490 &   -2.0601\\
$V_{(O2p-Mn3d)\pi}$        &   0.878961 &    1.0837\\
$V_{(Mn3d-Mn3d)\delta}$    &   0.066346 &    0.0488 \\ \hline 
\end{tabular}
\end{center}
\caption{Tight-binding parameters (in eV) for cubic LaMnO$_3$ and YMnO$_3$   
obtained by non-linear-least-squares fitting to the {\it ab initio}
eigenvalues along $\Gamma$ to X. E indicates an orbital energy,
and V an inter-atomic transfer integral. All transfer integrals
are between nearest neighbors, except those with the subscript `2'
which are between next nearest neighbors. Only the parameters listed 
in the table were allowed to be non-zero in the fitting procedure.}
\label{TB_params1}
\end{table}

\begin{table}[p]
\begin{center}
\begin{tabular}{r|c|c|}
      &  YMnO$_3$ & LaMnO$_3$     \\ \hline
FM    &     0      &     0         \\
A-AFM &   +140 meV  &  +110 meV     \\
PM    &  +0.70 eV  &  +1.16 eV     \\ \hline
\end{tabular}
\end{center}
\caption{Relative energies of different magnetic phases
in cubic paramagnetic YMnO$_3$ and LaMnO$_3$.}
\label{table1}
\end{table}


\begin{thebibliography}{99}

\bibitem{Jin}
S. Jin, T.H. Tiefel, M. McCormack, R.A. Fastnacht, R. Ramesh and L.H. Chen,
Science {\bf 264}, 413 (1994).

\bibitem{example}
See for example Proceedings of the 41st Annual Conference on Magnetism and
Magnetic Materials, Atlanta, GA, 1996 (in J. Appl. Phys. vol {\bf 81,8}
(1997)), and references therein.

\bibitem{Asamitsu}
A. Asamitsu, Y. Moritomo, Y. Tomioka, L. Arima and Y. Tokura,
Nature {\bf 373}, 407 (1995).

\bibitem{Kuwahara}
H. Kuwahara, Y. Tomioka, A. Asamitsu, Y. Moritomo and Y. Tokura,
Science {\bf 270}, 961 (1995).

\bibitem{Kuwahara2}
H. Kuwahara, Y. Tomioka, Y. Moritomo, A. Asamitsu, M. Kasai,
R. Kumai and Y. Tokura, Science {\bf 272}, 80 (1996).

\bibitem{Wood}
W.E. Wood, A.E. Austin, E.W. Collins and K.C. Brog, J. Phys. Chem.
Solids {\bf 34}, 859 (1973).

\bibitem{Schmid}
H. Schmid, Ferroelectrics {\bf 162}, 317 (1994).

\bibitem{HillJPC}
N.A. Hill, J. Phys. Chem. B., {\bf 104}, 6694-6709 (2000).

\bibitem{Waghmare_Rabe}
U.V. Waghmare and K.M. Rabe, Phys. Rev. B {\bf 55}, 6161 (1997).

\bibitem{Cohen_Krakauer}
R.E. Cohen and H. Krakauer, Ferroelectrics {\bf 136}, 95 (1992).

\bibitem{HillRabe}
N.A. Hill and K.M. Rabe, Phys. Rev. B. {\bf 59}, 8759 (1999)

\bibitem{Pickett_Singh}
W.E. Pickett and D.J. Singh, Phys. Rev. B {\bf 53}, 1146 (1996).

\bibitem{Yakel}
H.L. Yakel, W.C. Koehler, E.F. Bertaut and E.F. Forrat, Acta.
Cryst. {\bf 16} 957 (1963).

\bibitem{Bertaut1}
E.F. Bertaut, R. Pauthenet and M. Mercier, Phys. Lett. {\bf 7}, 110 (1963).

\bibitem{Bertaut2}
E.F. Bertaut, R. Pauthenet and M. Mercier, Phys. Lett. {\bf 18}, 13 (1965).

\bibitem{fiyi}
N. Fujimura, T. Ishida, T. Yoshimura, and T. Ito, App. Phys. Lett. {\bf 69}, 1011 (1996).

\bibitem{klk}
S. H. Kim, S. H. Lee, T. H. Kim, T. Zyung, Y. H. Jeong, and M. S. Jang, 
Cryst. Res. Technol. {\bf 35}, 19 (2000).

\bibitem{Huang}
Z.J. Huang, Y. Cao, Y.Y. Sun, Y.Y. Xue and C.W. Chu, Phys. Rev. B
{\bf 56}, 2623 (1997).

\bibitem{Frohlich}
D. Fr\"{o}hlich, St. Leute, V.V. Pavlov and R.V. Pisarev, Phys.
Rev. Lett. {\bf 81}, 3239 (1998).

\bibitem{Sa}
D. Sa, R. Valent\'{i} and C. Gros, cond-mat/9910283, accepted for
publication in European Journal of Physics B.

\bibitem{Quezel}
S. Quezel, J. Rossat-Mignod and E.F. Bertaut, Sol. Stat. Comm. {\bf 14},
941 (1974).

\bibitem{Brinks}
H.W. Brinks, H. Fjellv\o{a}g and A. Kjekshus, J. Sol. Stat. Chem.
{\bf 129}, 334 (1997).

\bibitem{Iliev} M.N. Iliev, M.V. Abrashev, H.-G. Lee, V.N. Popov,
Y.Y. Sun, C. Thomsen, R.L. Meng and C.W. Chu, Phys. Rev. B {\bf 57},
2872 (1998).

\bibitem{Yin}
M.T. Yin and M.L. Cohen, Phys. Rev. B {\bf 26}, 5668 (1982).

\bibitem{HKKS}
P. Hohenberg and W. Kohn, Phys. Rev. {\bf 136}, 864 (1964);\\
W. Kohn and L.J. Sham, Phys. Rev. {\bf 140}, 1133 (1965).

\bibitem{PWPPegs}
see for example T. Sasaki, A.M. Rappe and S.G. Louie, Phys. Rev. B 
{\bf 52}, 12760 (1995) (ferromagnetic Ni and Fe), S.P. Lewis, P.B. Allen 
and T. Sasaki, Phys. Rev. B {\bf 55}, 10253 (1997) (antiferromagnetic CrO$_2$)
and N.A. Hill and K.M. Rabe,
Phys. Rev. B. {\bf 59}, 8759 (1999) (perovskite manganites).

\bibitem{Rappe}
A.M. Rappe, K.M. Rabe, E. Kaxiras and J.D. Joannopolous,
Phys. Rev. B {\bf 41}, 1227 (1990).

\bibitem{Vanderbilt}
D. Vanderbilt, Phys. Rev. B {\bf 32}, 8412 (1985).

\bibitem{NLCCs}
S.G. Louie, S. Froyen and M.L. Cohen, Phys. Rev. B {\bf 26}, 1738 (1982).

\bibitem{Kleinman_Bylander}
L. Kleinman and D.M. Bylander, Phys. Rev. Lett. {\bf 48}, 1425 (1982).

\bibitem{Gonze}
X. Gonze, P. K{\" a}ckell and M. Scheffler, Phys. Rev. B {\bf 41},
12264 (1990).

\bibitem{Perdew_Zunger}
J.P. Perdew and A.Zunger, Phys. Rev. B {\bf 23}, 5048 (1981).

\bibitem{Ceperley_Alder}
D.M. Ceperley and B.J. Alder, Phys. Rev. Lett. {\bf 45}, 566 (1980).

\bibitem{vonBarth_Hedin}
U. von Barth and L. Hedin, J. Phys. C {\bf 5}, 1629 (1972).

\bibitem{Payne}
M.C. Payne, M.P. Teter, D.C. Allan, T.A. Arias and J.D. Joannopoulos,
Rev. Mod. Phys. {\bf 64}, 1045 (1992).

\bibitem{Castep}
M.C. Payne, X. Weng, B.Hammer, G. Francis, U. Bertram, A. de Vita,
J.S. Lin, V. Milman and A. Qteish, unpublished.

\bibitem{Monkhorst_Pack}
H.J. Monkhorst and J.D. Pack, Phys. Rev. B {\bf 13}, 5188 (1976).

\bibitem{Gilat-Raubenheimer}
G. Gilat and L.C. Raubenheimer, Phys. Rev. {\bf 144}, 390 (1966).

\bibitem{Alessio}
A. Filippetti and N.A. Hill, submitted to Phys. Rev. B

\bibitem{Mattheiss}
L.F. Mattheiss, Phys. Rev. B {\bf 6}, 4718 (1972).

\bibitem{Goodenough}
J.B. Goodenough, Phys. Rev. {\bf 100}, 564 (1955).

\end{thebibliography}
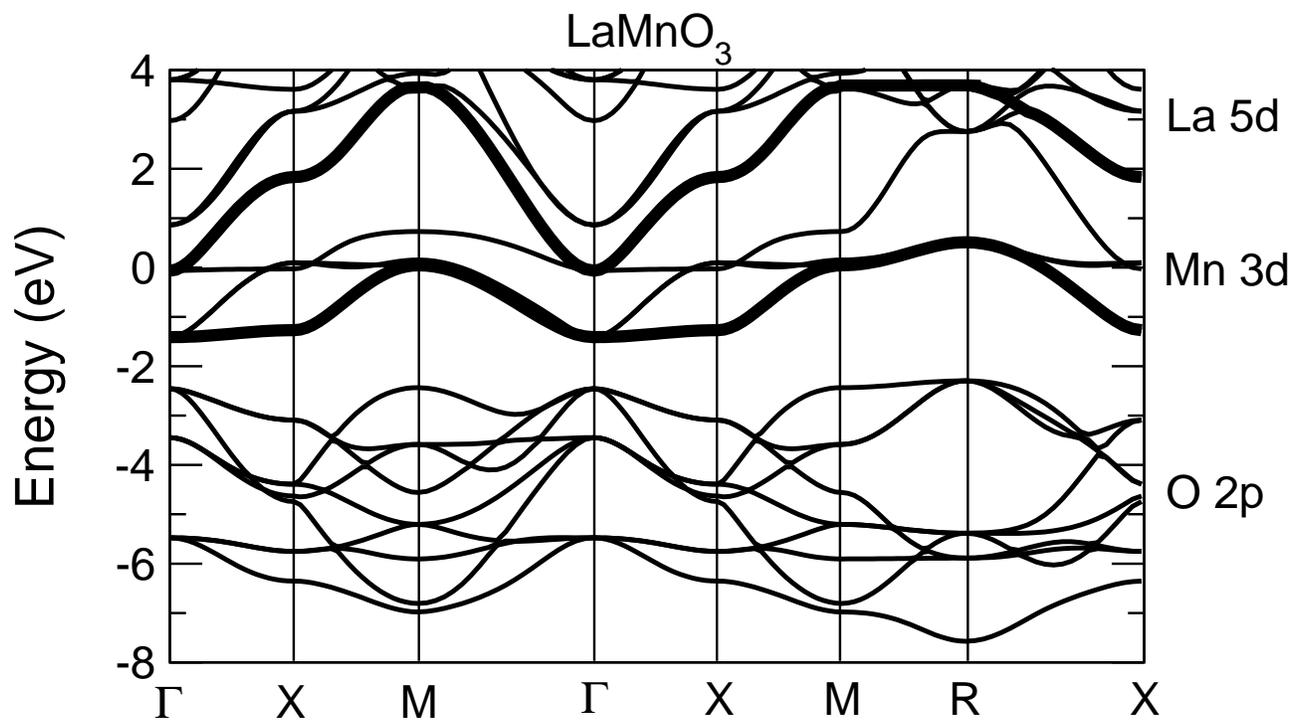
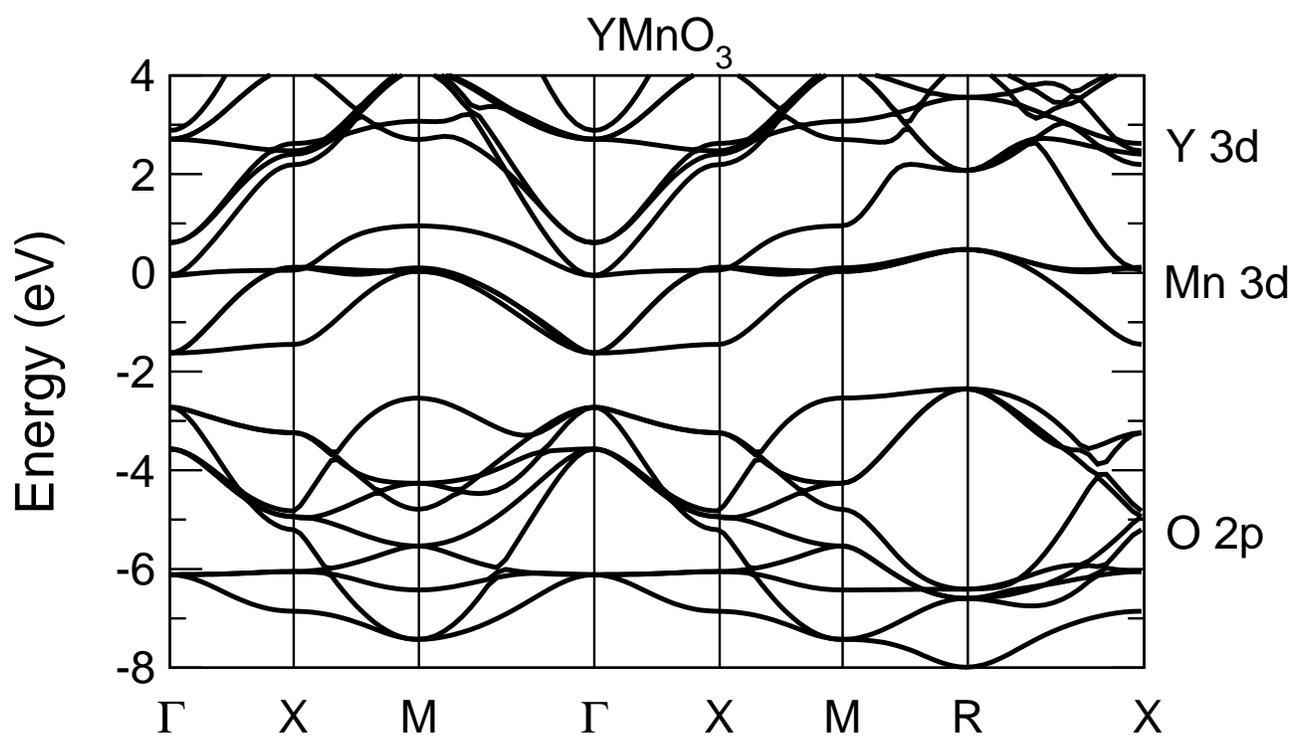
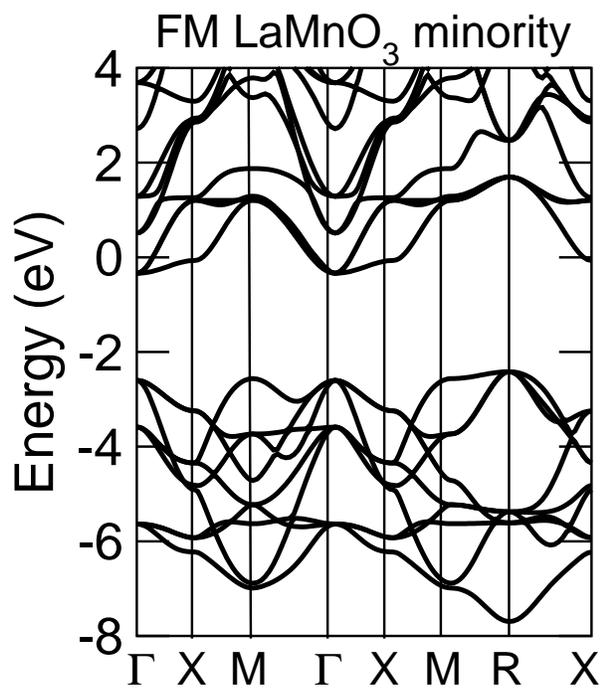
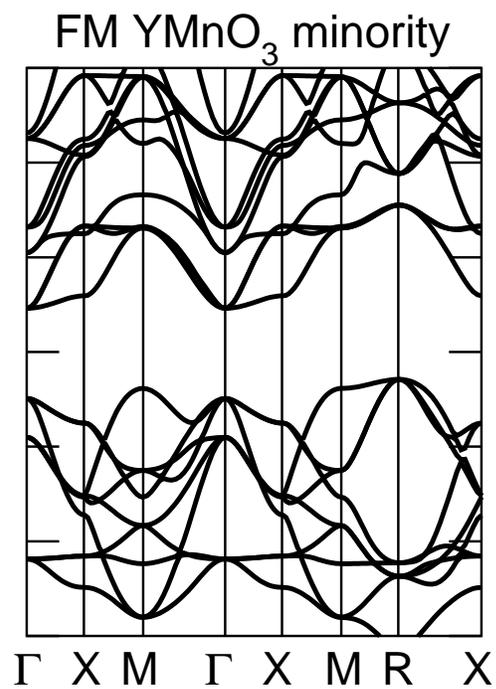
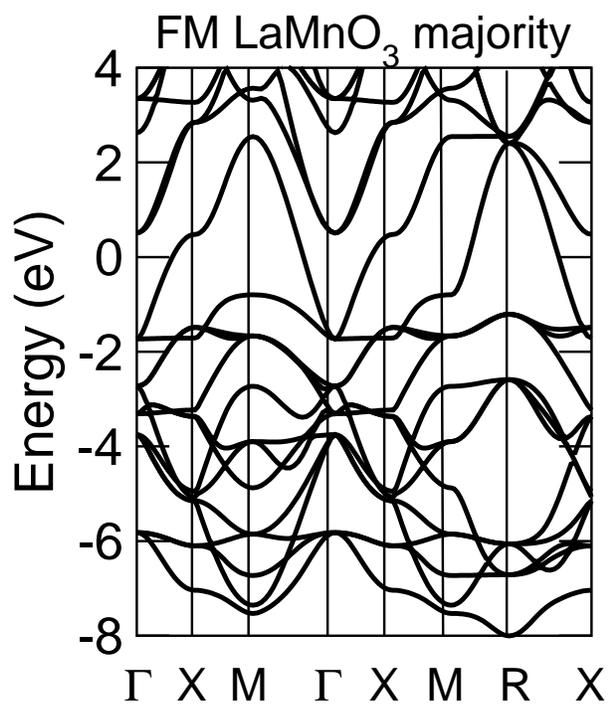
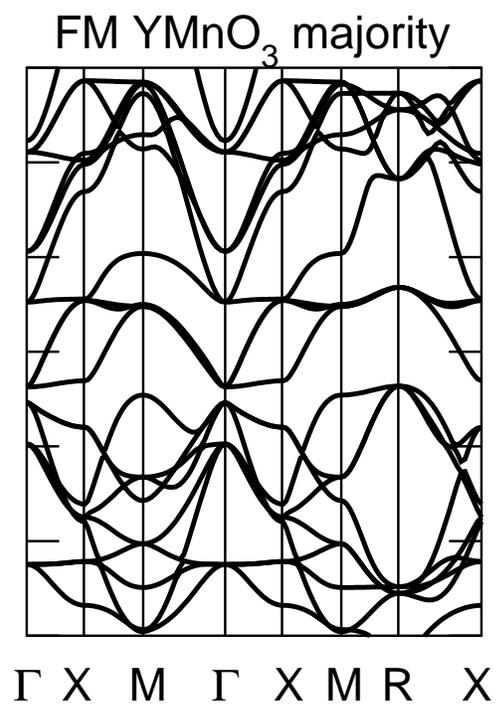
\end{document}